\documentclass[twocolumn,twoside,slac_two]{revtex4}
\usepackage{graphicx}
\usepackage{fancyhdr}
\begin{document}

\title{Physics of high energy atmospheric muons}

%

\author{Edgar Bugaev}
\affiliation{Institute for Nuclear Research, Russian Academy of Sciences,  \\
60th October Anniversary Prospect 7a, 117312 Moscow, Russia}

\begin{abstract}

In the first part of the talk the interesting new results of MINOS, OPERA and CMS
collaborations  (connected with the observational evidence of the rise in the muon
charge ratio at muon energies around 1 TeV) are briefly discussed from theoretical point of view.
A short review of charge asymmetric effects in muon energy losses is given.
In the second part of the talk, the modern theoretical approaches to the
problem of heavy quark production in high energy nucleon-nucleus interactions
are briefly considered (color dipole formalism, saturation models). The
recent new theoretical developments in the ancient problem of intrinsic
charm are also discussed.

\end{abstract}

\maketitle

\thispagestyle{fancy}


\section{THE CHARGE RATIO OF ATMOSPHERIC MUONS}

\subsection{Pika model}

A simple formula for the muon atmospheric charge ratio is obtained from the approximate
expression for the differential atmospheric muon spectrum \cite{ZK1960, Gaisser1990}:
\begin{equation}\label{f0}
\frac{d\phi_{\mu}}{dE_\mu} \cong \frac{d\phi_N}{dE_\mu} \sum_i
\frac{a_i Z_{Ni}} {1+b_i E_\mu \cos \theta^* / \epsilon_i} ,
\end{equation}
where $d\phi_N / dE_\mu \equiv d\phi_N/dE$ (at $E=E_\mu$)
is the primary spectrum of nucleons ($E_\mu$ is the muon energy
in atmosphere), index $i$ numerates muon parents ($\pi$, $K$ etc), the constants $a_i$, $b_i$
are purely kinematical (they depend on the characteristics of parent's decay into muons),
$ \epsilon_i$ are the critical energies, which are defined as parent's energies in atmosphere,
at which their interaction lengths are equal to decay lengths (for a vertical propagation),
$ \theta^*$ is the zenith angle of the muon trajectory (for simplicity, we assume that atmosphere
is flat). At last, $Z_{ij}$ is the spectrum weighted moment defined by the expression
\begin{equation}
\label{f2}
Z_{ij} = \int \limits_0^1 \frac{1}{\sigma_{ij}} \frac{d \sigma_{ij}}{dx_{lab}} x_{lab}^{\gamma-1} d x_{lab},
\end{equation}
where $\sigma_{ij}$ is the inclusive cross-section for the production of
a particle $j$ from the collision of a particle $i$ with a nucleus in the atmosphere and
$x_{lab}=E_j/E_i$ is the energy fraction carried by the secondary particle,
$\gamma$ is a spectral index of the primary nucleon spectrum, $\gamma=1.7$.

If we are interested in the interval of muon energies $(1 - 10^4)\;$GeV, one can take into
account only pions and kaons as muon's parents. In this case, the muon spectrum can
be expressed as
\begin{equation}\label{f1}
\frac{d\phi_{\mu^\pm}}{dE_\mu} \cong
\frac{a_\pi f_{\pi^\pm} } {1+b_\pi E_\mu \cos \theta^* / \epsilon_\pi} +
\frac{R_{K\pi} a_K f_{K^\pm}}{1+b_K E_\mu \cos \theta^* / \epsilon_K},
\end{equation}
where $R_{K\pi} = Z_{NK}/Z_{N\pi}$ and $f_{\pi^+}=1-f_{\pi^-}= Z_{N\pi^+}/Z_{N\pi}$,
$f_{K^+}=1-f_{K^-}= Z_{N K^+}/Z_{N K}$ are fractions of all pion and kaon decays
that yield positive muons. This simple formula used for an analysis of atmospheric muon charge
ratio data is a basic of the so-called ``pika model'' \cite{Schreiner:2009pf}. More exactly,
this model uses the following main assumptions:
1) particle ratios, $\pi^+/\pi^- = f_{\pi^+}$
and $K^+/K^- = f_{K^+}$, are energy independent; $K/\pi$ ratio is given by the
$R_{K\pi}\cdot a_K/a_\pi =0.149 \cdot 0.246/0.674 = 0.054$ \cite{Gaisser1990},
2) all the
spectrum weighted moments, $Z_{ij}$, are energy independent,
3) the muon charge ratio does not depend on the $E_\mu$ and the $\cos\theta^*$ separately, but
on the product  $E_\mu \cos\theta^*$, where $E_\mu$ is the energy of vertical muons on
the Earth surface,
4) contributions from charmed mesons are ignored.

\subsection{Particle ratios}

It is well known from studies of $K$-production in $NN$ and $NA$-collisions, in the energy region
near the threshold, that $K^+/K^-$ ratio is rather large. The explanation is simple: $K^+$-meson contains
$\bar s$-quark whereas, for a final baryon, $s$-quark is needed. The transfer of $s$-quark from
$K^-$-meson to a final baryon is realized most effectively through the reaction
$K^- N \to \pi \Lambda$. So, as a result, the escape of $K^-$-mesons is hampered.
Measurements of particle ratios at high energies, far from the threshold of $K$ production,
shows (see, e.g., the data of the BRAHMS Collaboration \cite{Bearden:2004ya} for $pp$-collisions
at $\sqrt{s}=200\;$GeV) that, at large rapidities (close to the fragmentation region), the
$K^+/K^-$-ratios are systematically larger than $\pi^+/\pi^-$-ratios. In general, such asymmetries
in the production of particles and antiparticles are indications of the leading particle effects
in hadronic collisions and are rather intensively studied now experimentally as well as
theoretically. In particular, large values of $K^+/K^-$-ratios measured by \cite{Bearden:2004ya}
are well explained by the DPM JET model \cite{Bopp:2005cr}.

\subsection{Muon energy losses: $z$-odd effects}

\subsubsection{Ionization energy losses} As is known, the standard expression
for energy loss contains, except of the main term, proportional to $z^2$, the tiny
term which is proportional to $\frac{\pi\alpha}{2} z^3$ ($\alpha$ is the fine structure
constant) \cite{Jackson:1972zz}. This term is due to the next-to-leading order
correction to the relativistic Rutherford formula. The $z^3$-correction to $dE/dx$
is taken into account in an analysis of data in the MINOS experiment
\cite{Schreiner:2009pf} as well as in the CMS \cite{Khachatryan:2010mw}, in spite
of its smallness. For $\mu^+$, the energy loss in matter is about $\sim 0.15\%$ higher
than for $\mu^-$, for CMS experiment, and it affects the muon charge ratio by less
than $\sim 0.3\%$ over the entire energy range \cite{Khachatryan:2010mw}.

\subsubsection{Radiative energy losses}

The muon bremsstrahlung cross section is given by the formula
\begin{equation}
\omega \frac{d\sigma}{d\omega} = \omega \left. \frac{d\sigma}{d\omega} \right|_{Born} +
\Delta_c + \Delta_{DBM} + \Delta_{nonleading}.
\end{equation}
Here, $\omega$ is the photon energy, the first term in r.h.s. is the cross section
calculated in the Born approximation (including, in this approximation, the
nuclear formfactor correction), $\Delta_c$ is the ``non-Coulomb'' correction
\cite{Andreev:1997pf} (it is important only for muons and is the non-Born part
of the correction which arises due to the non-Coulombic character of the electromagnetic
field of the nucleus having the extended size), $\Delta_{DBM}$ is the well-known
Coulomb correction of Davies, Bethe and Maximon \cite{Davies:1954zz}. At last,
$\Delta_{nonleading}$ takes into account the main correction to ultrarelativistic
approximation which had been used before for a calculation of all components of the
bremsstrahlung cross section.
It appears \cite{Lee:2004ina} that this correction is proportional to
$-\frac{m_\mu}{E_\mu} Z \alpha$, the minus sign corresponds to the $\mu^-$,
i.e., the $\mu^+$ has again a slightly larger energy loss. So, the
difference between the $\mu^+$ and $\mu^-$ energy losses
due to bremsstrahlung is inversely proportional to the muon $\gamma$-factor,
\begin{equation}
\left[\frac{dE}{dx}\right]^{\mu^+} - \left[\frac{dE}{dx}\right]^{\mu^-} \sim
\frac{Z \alpha}{\gamma} \langle \frac{dE}{dx} \rangle,
\end{equation}
and is very small for the case of the rock ($Z=11$). This difference is
negligible \cite{Schreiner:2009pf} in the region around $E_\mu\sim (0.1 - 1)\;$TeV,
where radiative energy losses become comparable with ionization energy losses.

\begin{figure}[t]
\includegraphics[width=9 cm, trim = 0 0 0 0]{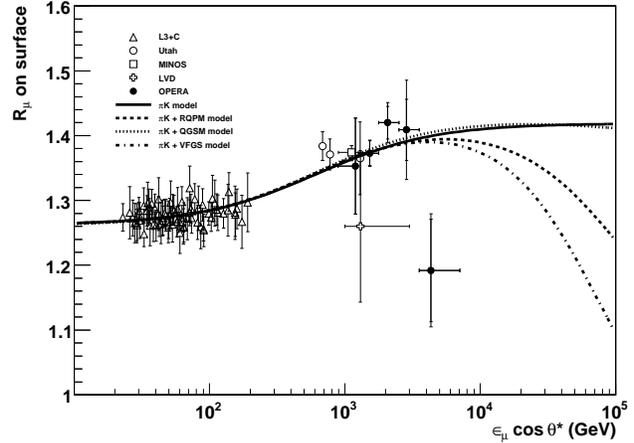}
\caption{$R_\mu$ values measured by OPERA, MINOS and other experiments
(taken from \cite{Agafonova:2010fg}). }
\label{fig-op}
\end{figure}

One should note that, of course, there are other sources of the charge asymmetry in muon energy losses.
For example, in the process of $e^+ e^-$-pair production by muon on atomic electrons such an
asymmetry arises due to interferences of different diagrams contributing to the
process \cite{Kelner:1998mh}. One should mention, in this connection, also the work
\cite{Kelner:1999yf} in which diffractive corrections to muon bremsstrahlung had been considered.

\subsection{Conclusions}

1. The recent experimental results of three groups, MINOS \cite{Schreiner:2009pf},
OPERA \cite{Agafonova:2010fg} and CMS \cite{Khachatryan:2010mw} (together with the old result of
Utah detector \cite{Ashley:1975uj}) show that the atmospheric muon charge ratio
at energies around 1 TeV and higher is sizeably larger ($R_\mu\sim 1.36 - 1.4$)
than at energies around $\sim 10^2\;$GeV (see Fig. \ref{fig-op}).

2. The data fittings give the similar results for all three groups:
$r_\pi = f_{\pi^+}/(1-f_{\pi^+}) = 1.241 \pm 0.035$,
$r_K = f_{K^+}/(1-f_{K^+}) = 2.26 \pm 0.29$ (MINOS N+F~\cite{Schreiner:2009pf});
$r_\pi = 1.229 \pm 0.001$, $r_K=2.12 \pm 0.03$ (OPERA~\cite{Agafonova:2010fg});
$f_{\pi^+} = 0.553 \pm 0.005$, $f_{K^+}=0.66 \pm 0.06$ (CMS~\cite{Khachatryan:2010mw}).

3. The increase of the ratio $R_\mu$ is due to the fact that
{\it i)} according to the pika model, at energies $\gtrsim 1\;$TeV, the contribution
of kaons in $R_\mu$ becomes dominant, and
{\it ii)} for kaons, the ratio $K^+/K^-$ is larger than the corresponding ratio for pions.

4. The fits of the pika formula to the old results of calculations \cite{Lipari1993}
and simulations \cite{Honda:1995hz, Fiorentini:2001wa} lead to the following numbers
\cite{Schreiner:2009pf}: $r_K=2.87$, $r_\pi=1.25$ (for the calculation
by Lipari \cite{Lipari1993}); $r_K=2.39$, $r_\pi=1.26$ (for the simulations with CORT
code \cite{Fiorentini:2001wa}); $r_K=1.63$, $r_\pi=1.28$ (for the simulations
by Honda \cite{Honda:1995hz}).

5. It would be quite interesting to check whether the recent experimental results
\cite{Bearden:2004ya} and theoretical predictions \cite{Bopp:2005cr} are consistent
with the new $R_\mu$-data.

\section{CHARM PRODUCTION IN $pA$-COLLISIONS AND PROMPT MUONS}

\subsection{Introduction}

One can see from Fig. \ref{fig-op} that the value of $R_\mu$ at $E_\mu \gtrsim 10\;$TeV strongly
depends on the charm production model. It is well known that it is not so difficult to calculate
muon atmospheric fluxes, but only if you know well the fluxes of muon parents (charmed mesons,
in a given case). So, the physics of hadronic charm production is intimately connected
with physics of atmospheric high energy muons.

\begin{figure}
\includegraphics[width=6.3 cm, trim = 0 0 0 0]{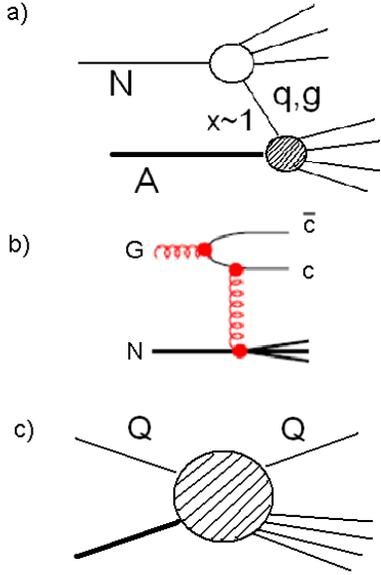}
\caption{The schematic diagrams of the $pA$-collision. \label{fig-diagr}}
\end{figure}

The schematic diagram of the process is shown in Fig \ref{fig-diagr}a.
We consider in this talk only two examples of such calculations.

1. If one uses perturbative QCD approach, the typical diagram for the lower blob
of diagram a) is shown in Fig. \ref{fig-diagr}b (gluon fusion).

2. If it is assumed that there are non-perturbative ``intrinsic'' heavy quark components in the
proton wave function \cite{Brodsky:1980pb}, the diagram in the lower blob of a) describes the
inclusive scattering of the heavy quark of the proton projectile on the target (Fig. \ref{fig-diagr}c).

The common points in these cases are following: the fragmentation region of the proton
projectile is of interest; the target is a nucleus; the proton energy is very
high ( $> 100 TeV$).
We will show in this talk that both cases can be studied using the modern approach
to high energy scattering in QCD based on the idea of the Color Glass
Condensate (CGC) (for a review, see \cite{Iancu:2003xm}).

\subsection{CGC framework}

\subsubsection{Dipole cross section}

In the CGC approach, the proton and the target nucleus are effectively described
as static random sources of color charge on the light cone. The leading
contribution to particle production in $pA$-collisions is obtained by calculating
the classical color field created by these sources and then averaging over
a random distribution \cite{McLerran:1993ni}.

The field is a solution of the classical Yang-Mills equation:
\begin{equation}
\left[ D_\mu , F^{\mu\nu} \right]  = J^\nu,
\end{equation}
where $J^\nu$ is the classical current which is determined by the sources $\rho_{p}$ and $\rho_A$:
\begin{equation}
J^\nu_a = g \delta^{\nu +} \delta(x^-) \rho_{p,a}(\vec x_\perp) +
 g \delta^{\nu -} \delta(x^+) \rho_{A,a}(\vec x_\perp).
\end{equation}
Here, $x^{\pm} = \frac{x^0 \pm x^3}{\sqrt{2}}$, $\vec x_\perp$ is the transverse
component of $\vec x$, $x^+$ is the ``LC time''.
$\rho_p, \rho_A$ are static, i.e., independent of the LC time.
$\delta(x^-)$ function assumes an infinitely thin sheet of color charge.

It is shown in many works that in CGC approach (see, e.g., \cite{Gelis:2002nn})
the amplitude for a scattering of a quark from the target, $qA\to qX$,
is expressed through the ``gauge rotation'' matrix $U(\vec x_\perp)$:
\begin{eqnarray}
{\cal M}_{qA\to qX} = \bar u(q)\times \;\;\;\;\;\;\;\;\;\;\;\;\;\;\;\;\;\;\;\;\;\;\;\;\;\;\;
\;\;\;\;\;\;\;\;\;\;\;\;\;\;\;\;\;\;\;
 \nonumber \\ \times \left[ \gamma^- \int d^2 x_{\perp} e^{i(q_\perp-p_\perp) x_\perp}
 \left( U(\vec x_\perp) - 1 \right) \right] u(p). \;\;\;
 \label{eq8}
\end{eqnarray}
This matrix is given by the expression:
\begin{equation}
U(\vec x_\perp) = T \exp \left[ -i g^2 \int \limits_{-\infty}^\infty dx^+
 \frac{1}{\Delta_\perp^2} \rho_{A,a}(x^+,x_\perp) t^a \right],
\end{equation}
$t^a$ is a color matrix, generator of the fundamental representation of $SU(N_c)$.
Here, $\rho_{A,a}(\vec x)$ is the density of color sources in the nucleus in the
covariant gauge  $\partial_\mu A^\mu = 0$. For concrete calculations the LC-gauge,
in which $A^+=0$, is used, because in this gauge the color fields depend only on the
matrix $U$ (i.e., they are ``pure gauge'' fields). But $U$ depends on the
$\rho_{A,a}(\vec x)$, the density of sources in the covariant gauge.

\begin{figure}
\includegraphics[width=8 cm, trim = 0 0 0 0]{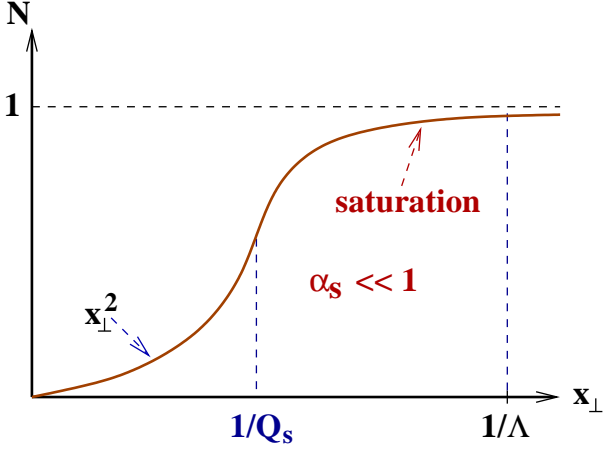}
\caption{The forward dipole amplitude, $\sigma_{\rm dip}/2\pi R^2$,
as a function of the transverse separation between $q$ and $\bar q$
(taken from \cite{JalilianMarian:2005jf}). \label{fig-satur} }
\end{figure}

It is very important that the same
matrix $U(\vec x_\perp)$ enters the expression for the dipole-hadron cross section,
$\sigma_{\rm dip}(\vec r_\perp)$, which is the total cross section
of interaction with nucleus, for a dipole of transverse size
$\vec r_\perp = \vec x_\perp - \vec y_\perp$ (with the quark located at $\vec x_\perp$
and the antiquark at $\vec y_\perp$ ) and is written as an integral over all the
impact parameters $\vec X_\perp = (\vec x_\perp + \vec y_\perp ) / 2$,
\begin{eqnarray}
\sigma_{\rm dip}(\vec r_\perp) \equiv \frac{2}{N_c} \int d^2 \vec X_\perp {\rm Tr_c}
 \langle
  1 - \;\;\;\;\;\;\;\;\;\;\;\;\;\;\;\;\;\;\;\;\; \nonumber \\
  - U \left( \vec X_\perp + \frac{\vec r_\perp}{2} \right)
   U^+ \left( \vec X_\perp - \frac{\vec r_\perp}{2} \right)
 \rangle _ \rho.
 \label{eqsdip}
\end{eqnarray}
We will see, in the next Section, that charm production cross sections contain
just the $\sigma_{\rm dip}(\vec r_\perp)$,
which is expressed, as is seen from Eq. (\ref{eqsdip}), through the 2-point function, or
``2-point correlator'', $\langle U(x_\perp) U^+(y_\perp) \rangle$.
In the Eq. (\ref{eqsdip}), ${\rm Tr}_c$ is the color trace ($U$ is the matrix in the color space),
and ensemble average is performed over the sources. In the McLerran-Venugopalan (M-V) \cite{McLerran:1993ni} model
a Gaussian distribution of color sources is used,
\begin{equation}
W[\rho] = \exp \left[ - \int dx^+ d^2 x_\perp
\frac{\rho_a(x^+, \vec x_\perp) \rho^a(x^+, \vec x_\perp)} {2 \mu_A^2 (x^+)}
\right],
\end{equation}
\begin{equation}
\langle {\cal O} \rangle \equiv \int [d\rho] W[\rho] {\cal O} [\rho].
\end{equation}
The color source is ``frozen'' through the collision due
to time dilation but changes from the collision to the subsequent one. Evidently,
$\mu_A^2 (x^+)$ is a density of color sources per unit volume and
\begin{equation}
\mu_A^2 = \int dx^+ \mu_A^2(x^+)
\end{equation}
is the average density of the sources per unit transverse area,
\begin{equation}
\langle \rho_a(\vec x_\perp) \rho_b(\vec y_\perp) \rangle =
 \delta_{ab} \delta(\vec x_\perp - \vec y_\perp) \mu_A^2,
 \;\;\; \mu_A^2 \sim A^{1/3}.
\end{equation}
The straightforward calculation gives for $\sigma_{\rm dip}(\vec r_\perp)$ the result
(if the target is homogeneous in the transverse plane):
\begin{equation}
\sigma_{\rm dip}(\vec r_\perp) = 2 \pi R^2 \left[ 1 - \exp \left( - Q_s^2 f(r_\perp) \right) \right],
\end{equation}
where $f(r_\perp)$ is the known function and $ Q_s^2$ is given by
\begin{equation}
Q_s^2 = \frac{g^4}{2} (t^a t_a) \mu_A^2 = \frac{g^4}{2} \frac{N_c^2 - 1}{2 N_c} \mu_A^2 \sim \alpha_s^2 A^{1/3}.
\end{equation}
By definition, $Q_s^2$ is the saturation scale (in the M-V model). The schematic form of
$\sigma_{\rm dip}(\vec r_\perp)$ is shown in Fig. \ref{fig-satur}.

Taking into account the quantum corrections \cite{Iancu:2003xm}, one can show that the dipole cross section
can be parameterized in the following form \cite{Kharzeev:2004yx}:
\begin{equation}
\sigma_{\rm dip}(r_\perp, x) \approx 2 \pi R_A^2 \left( 1 - \exp(-\frac{(Q_s^2(x) r_\perp^2)^\gamma }{4}) \right),
\end{equation}
\begin{equation}
\frac{Q_s^2(x)}{1 \; {\rm GeV}^2} = \frac{N_c^2 - 1}{2 N_c} \left( \frac{x_0}{x} \right)^\lambda A^{1/3},
\end{equation}
\begin{equation}
x_0 \approx 3 \times 10^{-4}, \; \lambda \approx 0.3, \; \gamma \approx 0.628.
\end{equation}
It follows from this equation that the saturation scale $Q_s^2$ becomes dependent
on energy, because the dipole cross section depend now on Bjorken $x$.
In DIS, the value of $x$ is connected with the photon virtuality,
\begin{equation}
x = \frac{Q^2}{2 M q_0}.
\end{equation}
In a case of the inclusive hadronic scattering and, in particular, in the case of heavy
quark production in hadronic collisions, $x$ is the momentum fraction of the
target parton participating in the reaction ($x=x_A$).

\subsubsection{Kinematics of reactions}

1. In the case of intrinsic charm production (Fig. \ref{fig-diagr}c), there is the simple connection
between $x_A$ and the momentum fraction of the projectile ($x_p$):
\begin{equation}
x_A = \frac{x_p q_t^2}{x_F^2 s} ; \;\; x_A \equiv x_p e^{-2 y_h},
\end{equation}
where $x_F$ is the Feynman $x$ of the produced particle; $y_h$ is its
rapidity, $q_t$ is its transverse momentum.
In the forward region one has
\begin{equation}
x_F \to 1 ; \;\;\; 1 \le x_p \le x_F, \;\; {\rm and } \;\;\;\; \langle x_A \rangle \to \frac{q_t^2}{s} \to 0.
\end{equation}

2. In the case of extrinsic charm production (Fig. \ref{fig-diagr}b) we have
two colliding gluons, with momentum fractions $x_g$ and $x_A$. Assume,
for simplicity, that the heavy quark pair is produced with Feynman $x_F$ corresponding to a fixed
$M_{Q\bar Q}$. One has
\begin{equation}
x_A = \left( \sqrt{x_F^2  + \frac{4 M_{Q\bar Q}^2}{s}} - x_F \right) \frac{1}{2},
\;\;\;\;\; x_g = x_F + x_A,
\end{equation}
and, again, in the forward region,
$x_A \sim M_{Q \bar Q}^2 / s  \to 0$.

\subsubsection{Target description in a small $x$ region}

1. Coherent lengths of small x partons in the target are very large,
$\ell \sim 1 / (M_N x)$, much larger than the nuclear size. Therefore, at very small
Bjorken $x$ inclusive particles are produced in hadron-nucleus collisions
coherently by the nuclear color field.

2. The gluon distribution which is the number of gluons per unit rapidity in the hadron
wave function,
\begin{equation}
x G(x, Q^2) = dN_{\rm gluons} / dy,
\end{equation}
grows with the decrease of $x$; as a result, the saturation scale $Q_s^2$ also grows:
\begin{equation}
Q_s^2 \sim A^{1/3} / x^\lambda.
\end{equation}
Physically, $Q_s^2$ is the average color charge squared of the gluons per
unit transverse area per unit rapidity.

3. If the density of color charge grows, the classical description of the color
field becomes possible as a first approximation \cite{McLerran:1993ni}. The nonlinear
effects in the classical Yang-Mills equations and quantum evolution corrections
lead to a saturation of gluon distributions.

One should note that the experimental as well as theoretical study of heavy quark
production at forward rapidities is very important just for the ``saturation physics''
because the saturation scale $Q_s$ increases with energy, $Q_s \sim A^{1/6} s^{\lambda/4}$,
and, at large $A$ and $s$, can be larger than $M_q$. The typical transverse momenta
of partons in the hadron are of the same order as $Q_s$ , so, it is natural to
expect that in such cases the probability of $QQ$-pair production will be greatly enhanced.

\subsection{Extrinsic and intrinsic charm production}

\begin{figure}[!b]
\includegraphics[width=8 cm]{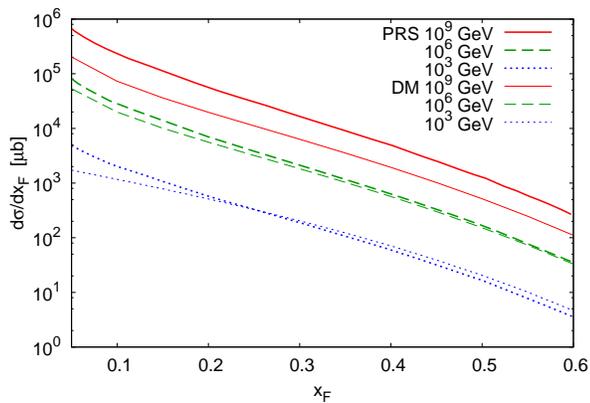}
\caption{The comparison of the predictions of the dipole model \cite{Enberg:2008te}
with the corresponding predictions of the pQCD model \cite{Pasquali:1998ji}, for different proton energies
(taken from \cite{Enberg:2008te}).}
\label{fig-Enberg}
\end{figure}

\begin{figure}[!b]
\includegraphics[width=7.8 cm]{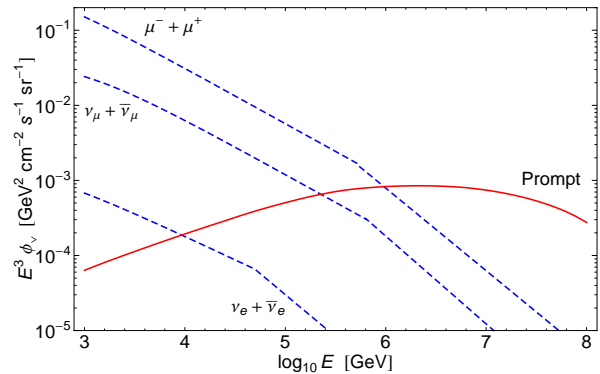}
\caption{The prompt muon flux predicted by dipole model \cite{Enberg:2008te}
($\nu_\mu + \tilde \nu_\mu$, $\nu_e + \tilde \nu_e$ and $\mu^+ + \mu^-$ - fluxes are approximately equal).
Conventional fluxes are represented by dashed lines \cite{Gondolo:1995fq}.
The figure is taken from~\cite{Enberg:2008te}.}
\label{fig-Enberg2}
\end{figure}

\subsubsection {Extrinsic charm in the color dipole approach}

In the model of \cite{Nikolaev:1995ty, Kopeliovich:2002yv} the differential cross section is
\begin{equation}
\frac{d\sigma}{dy}(pA \to Q\bar Q X) \cong x_g G(x_g,\mu^2) \sigma^{GA\to Q\bar Q X}(x_A,\mu^2),
\end{equation}
\begin{equation}
\sigma^{GA\to Q\bar Q X}(x_A,\mu^2) \cong \int dz d^2 r_\perp | \Psi_G |^2
 \sigma_{\rm dip}(x_A, r_\perp).
\end{equation}
Here, $| \Psi_G(z,\vec r_\perp) |^2$ is the probability of finding $QQ$-pair with a separation
$r_\perp$ and a fractional momentum $z$ in the gluon, $\mu \sim m_Q$ is  the factorization scale.
The cross section is written in a factorized form and contains $\sigma_{\rm dip}(x_A, r_\perp)$
which depends only on two-point correlator
$\langle U(x_\perp) U^+(y_\perp)\rangle$. It is a consequence of the special
choice of perturbative diagrams.
In general, if the nuclear color field is strong, one must take into account all orders in $\rho_A$
(if $g^2 \rho_A \sim 1$) and, in particular, include diagrams with exchange of two gluons
\cite{Blaizot:2004wv}.
After this there will be no simple factorization: the cross section will depend on
2-point, 3-point and 4-point correlators.

On Fig. \ref{fig-Enberg}, the predictions of this model \cite{Enberg:2008te} are shown, for
several incident proton energies, with a comparison with the corresponding results of pQCD
model \cite{Pasquali:1998ji}. One can clearly see the suppression of charm production at
very high energies due to gluon saturation effects. On the Fig. \ref{fig-Enberg2}, the
prediction of the same model \cite{Enberg:2008te} for the prompt lepton fluxes are shown
(in comparison with the conventional fluxes \cite{Gondolo:1995fq}).

\subsubsection {Intrinsic charm production.}

In the CGC approach, the diagram of the inclusive heavy quark scattering (Fig. \ref{fig-diagr}c)
is the basic one. The amplitude of this scattering is given by Eq. (\ref{eq8}).

After squaring this amplitude, one obtains, again, the formula containing 2-point correlator, which can be expressed through the dipole-hadron cross section $\sigma_{\rm dip}$. The formula for the inclusive
production of the charmed meson is \cite{Dumitru:2005gt, Goncalves:2008sw}:
\begin{eqnarray}
\frac{d \sigma (pA\to DX)}{dyd^2p_{\perp}} = \frac{1}{(2\pi)^2} \int\limits_{x_F}^{1} dx_p \frac{x_p}{x_F} f_{c/p}(x_p,Q^2) \times \nonumber \\
\times \int d^2 r_\perp e^{i\vec q_\perp \vec r_\perp} \frac{\sigma_{\rm dip}(r_\perp,x_A)}{2}
D_{D/c} \big(\frac{x_F}{x_p},Q^2 \big). \;\;\;
\label{dSigmadd}
\end{eqnarray}

The existence of an ``intrinsic charm'' (IC) component of the nucleon is not forbidden.
Many nonperturbative models predict such a component at an energy scale comparable to
mass of charm quark $m_c$. In \cite{Bugaev:2009jw}, the predictions of
the specific light-cone model developed in \cite{Brodsky:1980pb} (``BHPS model'') are
used for the calculation of charm production
(authors of \cite{Brodsky:1980pb} calculated the PDF of charm quarks from
the 5-quark component $uudcc$ in the proton). As a second example, authors of
\cite{Bugaev:2009jw} considered the phenomenological model
\cite{Pumplin:2007wg}, in which the shape of the charm PDF is sea-like, i.e.,
similar to that of the light flavor sea quarks (except for normalization).
In the concrete calculation, the scale dependence of PDFs must be taken into account.
For this aim, authors of \cite{Bugaev:2009jw} used the PDFs from the work \cite{Pumplin:2007wg},
based on the calculations of CTEQ group. For BHPS model, it was
assumed, in \cite{Bugaev:2009jw}, that the charm content of the proton is on the
maximal level, $\langle x \rangle _ {c+\bar c} = 0.02$. The same is for the sea-like IC model:
$\langle x \rangle _ {c+\bar c} = 0.024$.

\begin{figure}
\includegraphics[width=7 cm]{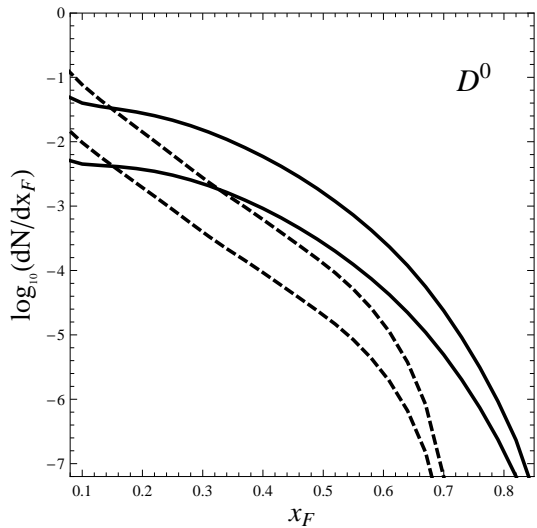}
\caption{The comparison of $x_F$-spectra for BHPS (solid) and sea-like
(dashed) models, for two energies, $\sqrt{s}= 10^2$ and $10^4$ GeV ($A=14$,
from \cite{Bugaev:2009jw}).\label{fig-our2}}
\end{figure}

\begin{figure}
\includegraphics[width=7 cm]{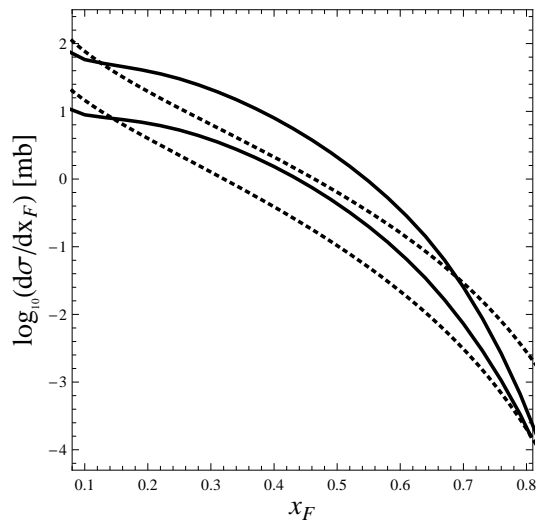}
\caption{The comparison of the results of \cite{Bugaev:2009jw} (thick) with the results of
\cite{Goncalves:2006ch} (thin), for $E_p = 10^9$ GeV (upper curves) and $3.4\times 10^4$ GeV
(lower curves); $A=14$. \label{fig-our4}}
\end{figure}

On the following two figures some results of the recent
work \cite{Bugaev:2009jw} are presented. Fig. \ref{fig-our2} gives the comparison of $x_F$-spectra
for two different models of the intrinsic charm. On Fig. \ref{fig-our4} the comparison
of the predictions for inclusive cross sections for intrinsic \cite{Bugaev:2009jw}
(the total contribution of all mesons) and extrinsic charm \cite{Goncalves:2006ch} is shown.
Authors of \cite{Goncalves:2006ch} used the dipole model approach and their results are similar
with the results of \cite{Enberg:2008te}.

\begin{figure}[!b]
\includegraphics[width=8.8 cm, trim = 0 40 0 0]{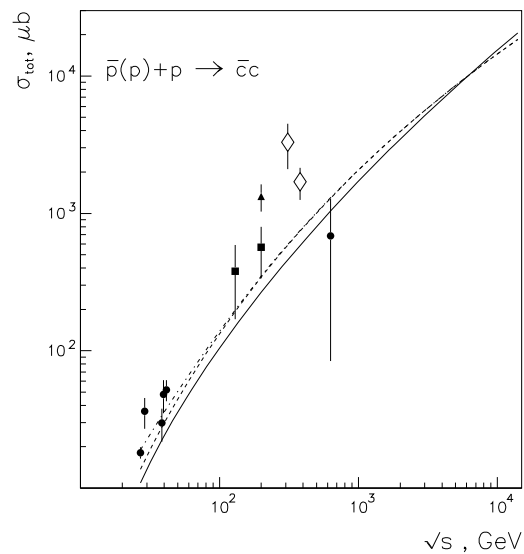}
\caption{The total cross section of charm production in $pp$ collisions
at $p>400 GeV/c$ \cite{Merino:2007rg}. At high energies, squares correspond to
PHENIX data and triangles to STAR data. The data of cosmic rays \cite{Rakobolslaya}
are shown by diamonds. The figure is taken from \cite{Merino:2007rg}.
\label{fig-rhic}}
\end{figure}

It follows from figure \ref{fig-our4} that in the tail region, $x_F > 0.7$, the curves
of \cite{Bugaev:2009jw} are even steeper
than those of \cite{Goncalves:2006ch}, in spite of their ``intrinsic'' origin. One must bear in
mind, however, that in \cite{Bugaev:2009jw} the inclusive spectra of charmed hadrons rather than $c\bar c$
pairs are calculated. Naturally, a taking into account of the fragmentation of $c$ quarks into
hadrons leads to additional steepening of $x_F$-spectra.

\subsubsection{Experimental data of RHIC}

One can see from the Fig. \ref{fig-rhic} that the results of different pQCD approaches
(presented on the Figure by curves) underestimate, in general, the cross section of charm
production at high energies. Authors of \cite{Merino:2007rg} argue, that it may be connected,
in particular, with large non-perturbative contributions in high density states (e.g.,
with CGC effects).

\subsection{Conclusions}

We tried to show in this review that the modern approach to studies of
proton-nucleus nucleus interactions based on the CGC theory can be effectively
used in calculations of heavy quark production in $pA$ – collisions.
The most important application of this approach is an use of it in calculations
of open charm production in collisions of cosmic rays in atmosphere (needed
for subsequent calculations of prompt muon and neutrino fluxes at energies around
100 TeV and higher). The reason is simple: in collisions of cosmic
rays in atmosphere, the fragmentation region of the projectile (or
forward rapidity region) is most important for calculations of all spectra
of secondary particles, and just in this region the effects of gluon saturation
studied in CGC theory are most efficient.





\end{document}